\documentclass[a4paper,11pt]{article}
\usepackage{pos}
\usepackage{physics}
\usepackage{dsfont}
\newcommand{\ee}{\mathrm{e}}
\newcommand{\ii}{\mathrm{i}}

\newcommand{\Vol}{\operatorname{Vol}}

\title{Toward the application of large-$N$
deconfinement to SU(3) QCD}
\author*[a]{Hiromasa Watanabe}
\affiliation[a]{
Yukawa Institute for Theoretical Physics, Kyoto University\\
Kitashirakawa Oiwakecho, Sakyo-ku, Kyoto 606-8502, Japan
}

\emailAdd{hiromasa.watanabe@yukawa.kyoto-u.ac.jp}
\abstract{
It is generally known for $\mathrm{U}(N)$ gauge theory at finite temperature that phase transitions are manifested by taking the large-$N$ limit. Since the large-$N$ theory undergoes two thermodynamic phase transitions, a nontrivial intermediate phase can be realized in addition to the phases classified as the conventional confined and deconfined phases. In this talk, we discuss that a similar picture can be applied to QCD with $N=3$. In particular, we analyze the gauge configurations of lattice QCD calculations involving dynamical quarks and show the results of an analysis of the deviation due to finite-temperature effects from the Haar randomness expected at zero temperature in $\mathrm{SU}(N)$ gauge theory, using physical pictures suggested by the large-$N$ theory.
}

\FullConference{Corfu Summer Institute 2023 "School and Workshops on Elementary Particle Physics and Gravity" (CORFU2023)\\
 23 April - 6 May, and 27 August - 1 October, 2023\\
Corfu, Greece\\}

\begin{document}
\maketitle

%-------
\section{Introduction}
%-------
The confinement/deconfinement phase transition in gauge theories~\cite{Polyakov:1978vu,Susskind:1979up} provides an important comprehension of a variety of fields in physics beyond its academic interests, from the quark matters in Quantum Chromodynamics (QCD) under extreme circumstances to the black hole geometries via the gauge/gravity duality.
In most cases, the strong coupling nature of Quantum Field Theory (QFT) prevents us from approaching the intriguing phenomenon. Lattice QCD based on the Monte Carlo simulation has been established and recognized as a powerful tool for tackling this difficulty. 
Various nonperturbative aspects of QFTs have been investigated in the framework until now, and the attempt to unveil the QCD phase structure is continuing. 
Remarkably, it is expected that the QCD thermal phase transition is not an actual transition but a crossover for light quark mass~\cite{Aoki:2006we}, while it is the first-order transition for heavy quarks and the pure Yang-Mills theory.

Another effective approach to the phenomenon is to take an idealized limit, which allows us to analyze theory more easily. 
Investigation of large-$N$ gauge theories, categorized in this approach, has been successful in exploring the phase transition kinematically since the saddle-point approximation of the path integral becomes exact and the free energy evaluated by the saddle-point configurations plays a role in the order parameter for the thermal transition. 
More specifically, the scaling of the free energy with respect to the rank of gauge group $N$ captures the phase transition, $F = O(N^0)$ for the confined phase and $F = O(N^\alpha)$ for the deconfined phase with nonzero $\alpha$.
Moreover, the Polyakov line (i.e., the holonomy matrix whose trace gives the Polyakov loop) can distinguish the phases. 
It has been recognized that the Polyakov line refers to the gauge symmetry rather than the center symmetry, as explained in Sec.~\ref{sec:deconf_largeN}.

\begin{figure}[t]
    \centering
    \scalebox{0.25}{\includegraphics{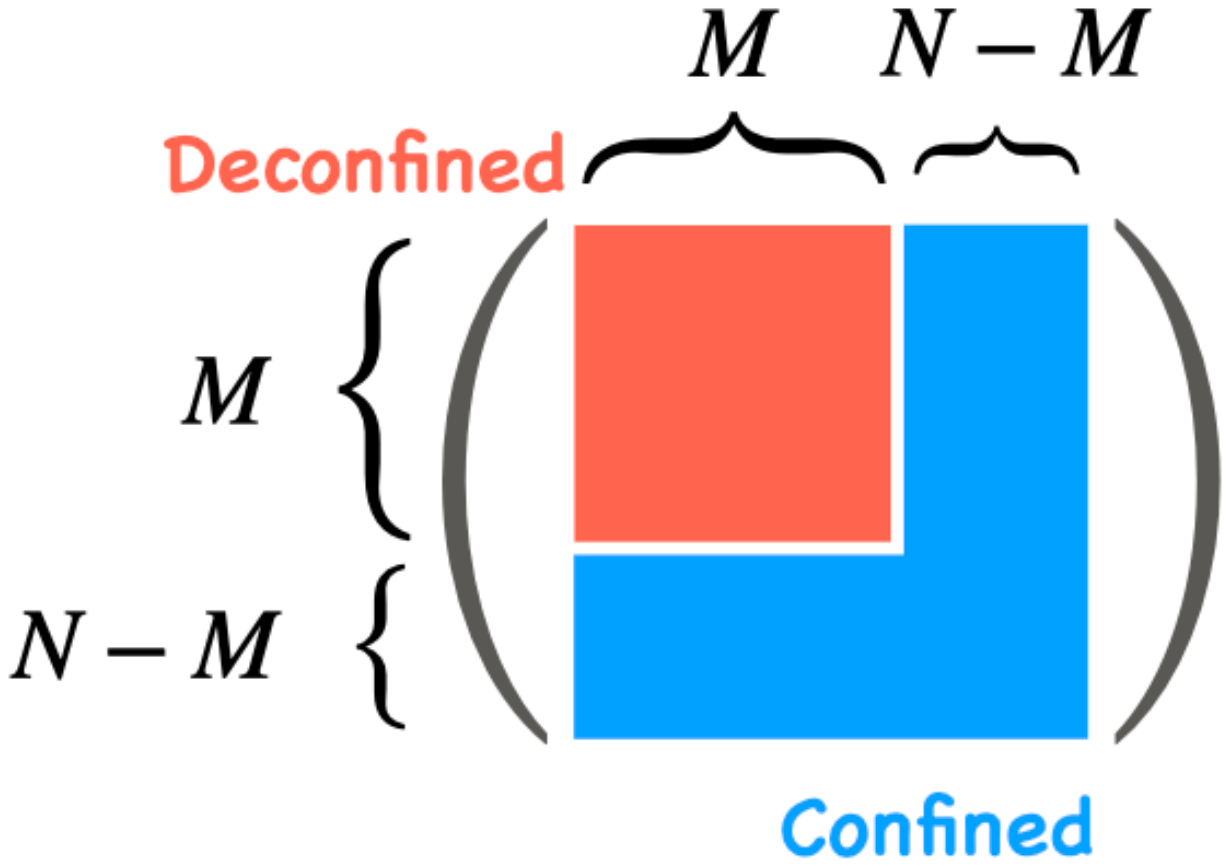}}
    \scalebox{0.25}{
    \includegraphics{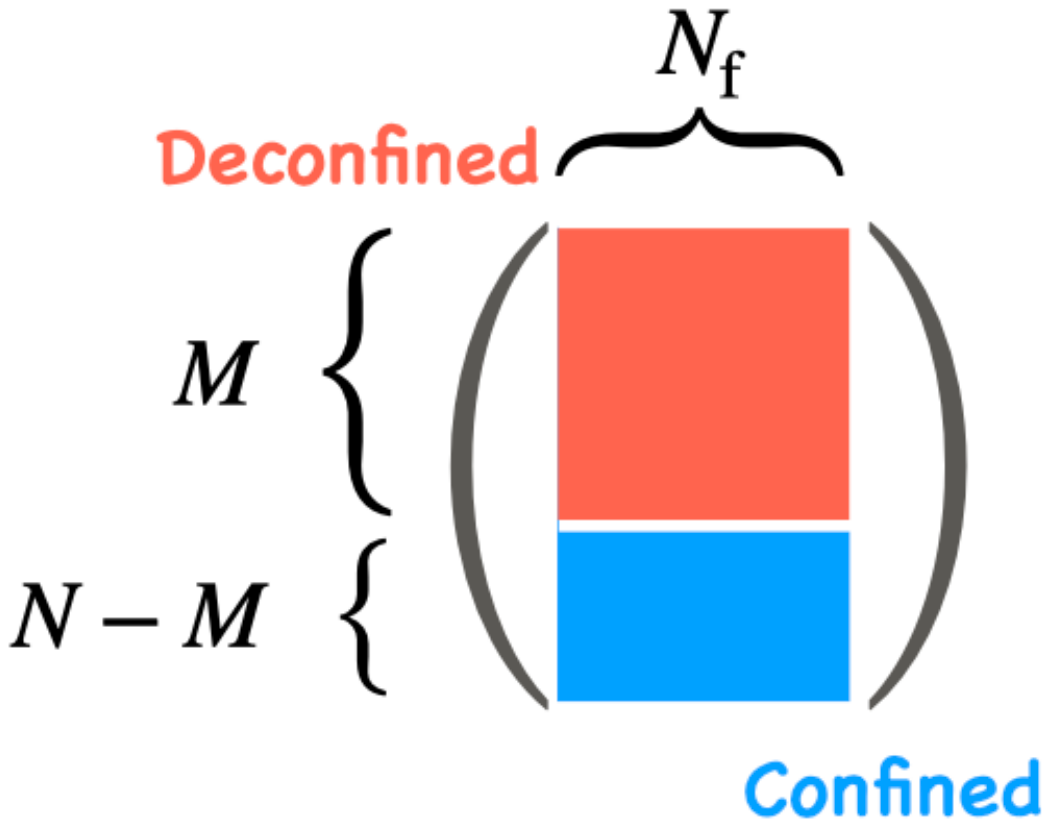}}
    \caption{Schematic pictures of the representative field configuration in the partially-(de)confined phase. [Left] Gauge and adjoint matter fields. Upper $M\times M$ submatrix only deconfines. [Right] Fundamental matter field with $N_\mathrm{f}$ flavors. Depending on the setup, two-phase coexistence in the flavor sector might happen, as discussed in Ref.~\cite{Hanada:2019kue}. Note that any choice of embedding is connected through the gauge transformation, and the gauge-transformed configurations of these are equivalent to the above.}
    \label{fig:partial_deconf}
\end{figure}

It is also convenient to apprehend the criterion for deconfinement by free energy in a more intuitive way. 
In the confined phase, since any color degrees of freedom are condensed, the free energy (and thermodynamic quantities deriving from it) scales as $N^0$, up to zero-point contributions. In the deconfined phase, it scales as $N^2$ for the case that the gauge field is in the adjoint representation, which means that all color degrees of freedom are excited and contribute to thermodynamics.
On the other hand, a question arises: what happens for the intermediate energy or temperature region, for example, if the energy is set to $\epsilon N^2$ and $\epsilon$ is of order $N^0$? In such cases, what we call \textit{partial deconfinement} takes place. 
Roughly speaking, a number of color degrees of freedom inside the $M\times M$ subsector $M\sim\sqrt{\epsilon}N$ can only be excited by the limited energy (see Fig.~\ref{fig:partial_deconf}).

Although the large-$N$ limit is often a mere prescription for idealization, it has an essential role physically in certain setups, such as the gauge/gravity duality in the context of quantum gravity. 
Interestingly, the AdS/CFT correspondence~\cite{Maldacena:1997re,Gubser:1998bc,Witten:1998qj,Aharony:1999ti}, which is the most well-established holographic duality in the superstring theory, has predicted the equivalence between the deconfinement transition and the Hawking-Page transition~\cite{Hawking:1982dh} that separates the geometry with and without a black hole and a mysterious intermediate phase dual to the so-called small black hole.
In retrospect, there are seminal papers~\cite{Sundborg:1999ue,Aharony:2003sx} exploring the dual QFT side at weak coupling. 
They found the presence of two distinct phase transitions at strict large $N$, the Hagedorn transition~\cite{Hagedorn:1965st} and Gross-Witten-Wadia (GWW) transition~\cite{Gross:1980he,Wadia:2012fr}.
They also conjectured that, if one considers the strong coupling regime, the region distinguished by the two strict large-$N$ transitions will give a dual description of the above intermediate phase.
The basic idea of partial deconfinement was originally introduced~\cite{Hanada:2016pwv} in this context. Later, Refs.~\cite{Berenstein:2018lrm,Hanada:2018zxn} and subsequent research pointed out that partial deconfinement mentioned above can take place more generically in various large-$N$ theories.

The successive question to be addressed is how the above concepts can be generalized to finite-$N$ theories at finite temperature and whether the analogous phase structure is really possible to realize.
As a first step, we take a rather phenomenological approach, namely, perform numerical analyses on the lattice QCD gauge configurations that contain the dynamical effect of quarks.
By the use of configurations generated by the WHOT-QCD collaboration~\cite{Umeda:2012er}, we compute several observables related to the Polyakov loop and aim to characterize the phases by their expectation values 
 and scaling behaviors.

This conference proceeding is organized as follows. 
In Sec.~\ref{sec:deconf_largeN}, we overview the deconfinement in the large-$N$ gauge theories. 
We discuss that, in the intermediate energy region, the two-phase coexistence in terms of the color degrees of freedom can be seen and the mechanism occurring it is essentially the same as the Bose-Einstein condensation.
A nontrivial relation to the black hole geometry in the context of superstring theory is also commented on.
The application of the notion of large-$N$ deconfinement to finite-$N$ theory, in particular, SU(3) QCD with dynamical fermion is elaborated in Sec.~\ref{sec:QCD}, following recent works~\cite{Hanada:2023krw-ptep,Hanada:2023rlk-ptep}.
Sec.~\ref{sec:conclusion} is devoted to the conclusion and discussion.

%-------
\section{Large-$N$ deconfinement}
\label{sec:deconf_largeN}
%-------
In this section, we will overview the partial (de)confinement from the perspective of a description by the eigenphase of the Polyakov line and the underlying mechanism.
%-------
\subsection{A description by Polyakov line}
%-------
Let us begin with the criterion of the confinement/deconfinement phase transition in gauge theories.
The Polyakov line plays a crucial role, defined by the gauge field along the temporal direction as 
\begin{equation}
    P = \operatorname{P}\exp\qty(\ii\oint A_t),
\end{equation}
where $\operatorname{P}$ represents the path-ordered multiplication.
For the theories with U($N$) or SU($N$) gauge group, this is an $N\times N$ unitary matrix.
Then, we can introduce the Polyakov loops with winding $n$ as
\begin{equation}
    u_n = \frac{1}{N} \Tr P^n.
\end{equation}
Intuitively, these quantities detect the free energy contribution of the probe degrees of freedom, and $u_n \ne 0$ for some $n$ indicates the phase transition to the deconfined phase.
For a specific case, this transition is called the Hagedorn transition at which the long excited string starts to favor.

\begin{figure}[t]
    \centering
    \includegraphics[width=\textwidth]{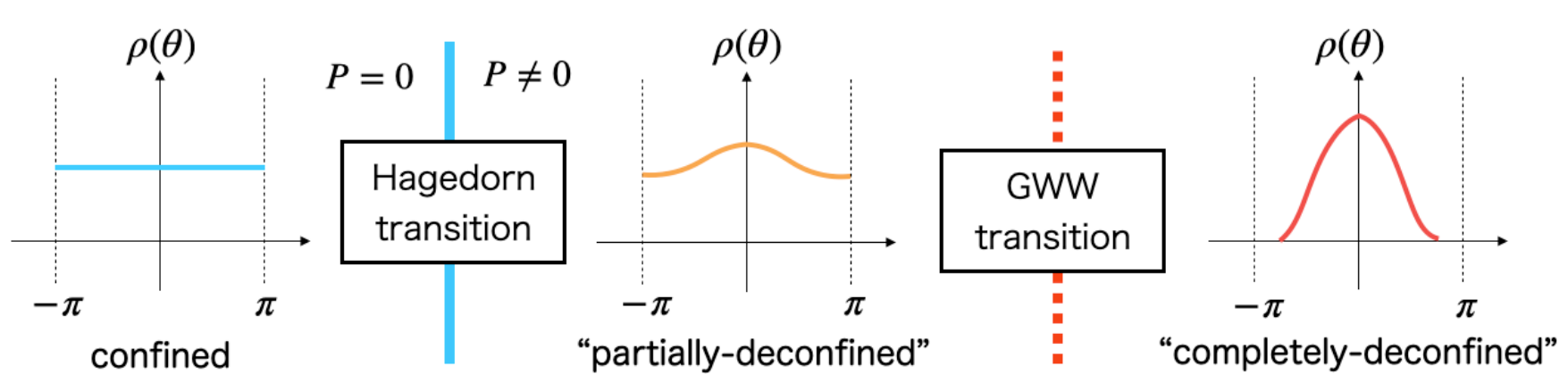}
    \caption{Schematic picture for the distribution of the Polyakov line phases $\rho(\theta)$.}
    \label{fig:phase-dist}
\end{figure}

Furthermore, let $\theta_j$ and $\rho(\theta)$ be the eigenphase of $P$ and its distribution, normalized as $\int_{-\pi}^{\pi} \dd\theta\rho(\theta)=1$, respectively.
Since $u_n = \frac{1}{N}\sum_j \ee^{\ii n\theta_j}$, the phase distribution can be expressed by the Fourier expansion as
\begin{equation}
    \rho(\theta) 
    =
    \frac{1}{N}\sum_{j=1}^N\delta(\theta-\theta_j)
    =
    \frac{1}{2\pi}\sum_{n=-\infty}^{\infty} u_n \ee^{-\ii n\theta},
    \label{eq:rho_Fourier_exp}
\end{equation}
which forms a continuous distribution between $-\pi$ and $+\pi$ in the large-$N$ limit. 
The $\mathbb{Z}_N$ center shifts $\theta$ constantly, and therefore, the confined phase corresponds to the uniform distribution $\rho(\theta)=\frac{1}{2\pi}$, while the deconfined phase is characterized by non-uniform distribution. 
(Fig.~\ref{fig:phase-dist}.)
It has been known~\cite{Aharony:2003sx,Sundborg:1999ue} that the non-uniform distributions can be further distinguished by the Gross-Witten-Wadia (GWW) transition that specifies a finite interval obeying $\rho(\theta)=0$ for $\abs{\theta} > \theta_0$.
See the right panel of Fig.~\ref{fig:phase-dist}.
The distribution with a finite interval can be realized only when all Fourier modes $\qty{u_n}$ have nonzero values.
This is a characteristic behavior of the phase above the GWW transition point.
It approaches the delta function at the high-temperature limit, reflecting that all multiply-wound loops contribute equally.

Recently, it was pointed~\cite{Hanada:2016pwv,Berenstein:2018lrm,Hanada:2018zxn,Hanada:2019czd} out that the intermediate phase exhibits the two-phase coexistence of the confined and deconfined sectors in terms of the color degrees of freedom, which is referred to as \textit{partial (de)confinement}, as depicted in Fig.~\ref{fig:partial_deconf}.
Representatively, it turns out that, for certain cases~\cite{Sundborg:1999ue,Aharony:2003sx,Schnitzer:2004qt} such as 4d weakly-coupled Yang-Mills theory, the distribution function of the intermediate phase can be expressed by the ones defined at the two transition points as
\begin{equation}
    \rho(\theta,\beta) 
    = 
    \qty(1-\frac{M}{N})\rho_\mathrm{con}(\theta) + \frac{M}{N}\rho_\mathrm{GWW}(\theta,M)
    =
    \frac{1}{2\pi}\qty(1-\frac{M}{N}) + \frac{M}{N}\rho_\mathrm{GWW}(\theta,M),
\end{equation}
where the size of the deconfined sector $M$ is determined thermodynamically, and practically, it can be read by the Polyakov loop $P=u_1$. 
For the above specific cases, the free energy is completely described by the Polyakov loops $\qty{u_n}$, and equivalently, the distribution function $\rho(\theta)$, and we demonstrated that thermodynamic quantities consist of the contributions from the confined and deconfined sectors with the ratio $M/N$~\cite{Hanada:2018zxn,Hanada:2019czd,Hanada:2019kue}.
It was discussed in Ref.~\cite{Watanabe:2020ufk} that the separation can take place beyond the weak coupling regime.

Since the center symmetry does not distinguish the phases, the Polyakov loops have a certain meaning independent of the center symmetry. 
This is also supported by the fact that the Polyakov loop can characterize the phases in $N_\mathrm{f}$-flavor large-$N$ QCD with $N_\mathrm{f}/N$ fixed at weak coupling~\cite{Schnitzer:2004qt} that does not enjoy explicit center symmetry due to quarks in the fundamental representation.

%-------
\subsection{Mechanism of large-$N$ deconfinement}
\label{sec:mechanism}
%-------
The mechanism of how the two-phase coexistence in color space takes place thermodynamically is clarified from the viewpoint of symmetry in the extended Hilbert space~\cite{Hanada:2020uvt}.
A crucial point is to change our perspective from the gauge-invariant Hilbert space $\mathcal{H}_\mathrm{inv}$ to the extended one $\mathcal{H}_\mathrm{ext}$.
Even when we deal with nonsinglet states directly from the latter perspective, the correct partition function can be computed 
\begin{equation}
    Z(\beta) 
    =
    \frac{1}{\Vol G}\int_G \dd g \Tr_{\mathcal{H}_\mathrm{ext}}\qty(\hat{g}\,\ee^{-\beta \hat{H}})
    =
    \Tr_{\mathcal{H}_\mathrm{inv}}\qty(\ee^{-\beta \hat{H}}),
\end{equation}
by projecting states into the gauge-invariant sector. 
The integral is performed over $g\in G$ with the Haar measure, and the operator $\hat{g}$ is the representation of $g$ acting on $\mathcal{H}_\mathrm{ext}$.

As an example, we take the energy eigenstate $\ket{\phi}$, which is nonsinglet in general.
Then, the matter to answer the above question is how much $\int_G \dd g \mel{\phi}{\hat{g}}{\phi}$, in other words, how much the states overlap before and after acting $\hat{g}$.
If this becomes large, it means that corresponding gauge-equivalent states give a nonnegligible contribution to thermodynamics.
In particular, the confining vacuum yields a significant enhancement at low energies, while the deconfined states do not so much. 
It is important to notice that $\hat{g}$ is nothing but the Polyakov line and the amount of enhancement essentially specifies the size of the deconfined sector $M$.
The constant distribution for the completely-confined phase is certainly translated into the Haar-random distribution of ``SU($\infty$),'' reflecting the symmetry of vacuum $\hat{g}\ket{\mathrm{vac.}} = \ket{\mathrm{vac.}}$.
In this sense, the mechanism provides a precise interpretation of the two-phase coexistence in the color space illustrated in Fig.~\ref{fig:partial_deconf}.

This mechanism can also be applied to the large-$N$ QFTs~\cite{Hanada:2023rlk-ptep}. 
Let us consider (3+1)d SU($N$) lattice Yang-Mills theory on the lattice. We employ the Kogut-Susskind Hamiltonian formalism~\cite{Kogut:1974ag}, and only the temporal direction is continuous.
Aiming to the continuum limit, we focus on the weak coupling,\footnote{
The strong coupling limit is also a qualitatively intriguing setup to investigate the confinement/deconfinement transition, while it is contaminated by lattice artifacts. See Refs.~\cite{Gautam:2022exf,Hanada:2023rlk-ptep}.
} and the Hamiltonian can be described by the magnetic part as
\begin{equation}
    \hat{H}_{\rm B}
    =
    -\frac{1}{2ag^2}\sum_{\vec{n}}\sum_{\mu<\nu}
    \left(
    \hat{U}_{\mu,\vec{n}}
    \hat{U}_{\nu,\vec{n}+\hat{\mu}}
    \hat{U}^\dagger_{\mu,\vec{n}+\hat{\nu}}
    \hat{U}^\dagger_{\nu,\vec{n}}
    +
    \mathrm{h.c.}
    \right),
\end{equation}
where $a$ is the lattice spacing for the spatial extent, and $\hat{U}_{\mu,\vec{n}}$ is an operator that corresponds to the link variable $U_{\mu,\vec{n}} \sim \ee^{\ii ag A_\mu(\vec{n})}$ of the group manifold SU($N$).

The extended Hilbert space can be spanned as 
${\cal H}_{\rm ext}
=
\otimes_{\mu,\vec{n}}{\cal H}_{\mu,\vec{n}}
\sim
\otimes_{\mu,\vec{n}}
\left(
\oplus_{g\in{\rm SU}(N)}
|g\rangle_{\mu,\vec{n}}
\right), 
$
with states $\ket{g}_{\mu,\vec{n}}$ such that $
\hat{U}_{\mu,\vec{n}}
\ket{g}_{\mu,\vec{n}}
=
g\ket{g}_{\mu,\vec{n}}\,
$, and $g \in {\rm SU}(N)$.
The symmetry responsible for the enhancement in the partition function is the now $G = \prod_{\vec{n}}\qty[{\rm SU}(N)]_{\vec{n}}$, and the enhancement is expected around the configuration $U_{\mu,\vec{n}} \sim \mathds{1}$.
This is however incomplete since that configuration is affected by the gauge transformation $\Omega_{\vec{n}}\in {\rm SU}(N)$.
To take the gauge transformation into account, let us consider the pure-gauge configuration 
$
U_{\mu,\vec{n}}
    =
\Omega_{\vec{n}}^{-1}\Omega_{\vec{n}+\hat{\mu}}
$
.
For a ``local'' gauge transformation $\Omega'_{\vec{n}}:=\Omega_{\vec{n}}^{-1}V_{\vec{n}}\Omega_{\vec{n}}$, 
it leads 
\begin{equation}
    U_{\mu,\vec{n}}
    =
    \Omega_{\vec{n}}^{-1}\Omega_{\vec{n}+\hat{\mu}}  
    \longrightarrow
    \Omega_{\vec{n}}^{\prime -1}    \qty(\Omega_{\vec{n}}^{-1}\Omega_{\vec{n}+\hat{\mu}})\Omega^{\prime}_{\vec{n}+\hat{\mu}}
    =
    \Omega_{\vec{n}}^{-1}
    \qty(
    V_{\vec{n}}^{-1}
    V_{\vec{n}+\hat{\mu}}
    )
    \Omega_{\vec{n}+\hat{\mu}},
\end{equation}
and it gives an enhancement if the $V_{\vec{n}}$ is independent of $\vec{n}$ or \textit{slowly-varying} i.e., $V_{\vec{n}}^{-1} V_{\vec{n}+\hat{\mu}}$ is close to $\mathds{1}$.
Note that the meaning of the closeness is somehow obscure unless a suitable renormalization prescription is given. We will discuss the related issue in Sec.~\ref{sec:conclusion}.

In the partially-(de)confined states, a smaller but huge enhancement factor can arise from the confined sector invariant under transformations of the subgroup of the original gauge group $G$. This will be observed at the configuration level for $V_{\vec{n}}$ or the plaquette, as illustrated in Fig.~\ref{fig:partial_deconf}.
The lesson of this observation is that \textit{the eigenphase distribution of the Polyakov line must be Haar-random} for confining vacuum.
This indicates a stronger condition than the unbroken center symmetry.
As discussed in Sec.~\ref{sec:QCD}, the analogous enhancement can appear even for the finite-$N$ theories, including QCD.

%-------
\subsection{Relation to quantum gravity}
%-------
Hints for the existence of partially-(de)confined phase were pointed out a long time ago in the context of quantum gravity, specifically, string theory.
Focusing on two different aspects of the theory, it gradually turned out that there is equivalence between a pair of theories with and without gravitational interaction.
The equivalence is often called gauge/gravity duality, and AdS/CFT correspondence~\cite{Maldacena:1997re,Gubser:1998bc,Witten:1998qj,Aharony:1999ti} is one of the most successful and established examples of it.\footnote{
For other cases of dualities, see, for example, Ref.~\cite{Hanada:2022wcq} and references therein.
}
On the gravity side, 10d AdS-Schwarzschild black hole geometry (sometimes called the large or big black hole) is thermodynamically favored at high energies. In contrast, the thermal AdS geometry (i.e., graviton or string gas phase) is realized at low energies, and these phases are separated by the Hawking-Page transition~\cite{Hawking:1982dh} which is a first-order transition in the canonical ensemble. Namely, these phases correspond to the local minima of free energy and the global minimum changes depending on the temperature.
The duality conjectures the same phase structures for both dual theories, and the transition is interpreted as the confinement/deconfinement transition on the dual gauge theory side.

More strictly speaking, the intermediate energy region is also thermodynamically stable~\cite{Horowitz:1999uv} and realized as a physical phase in the microcanonical ensemble~\cite{Aharony:1999ti}.
In that region, the so-called small black hole and Hagedorn string appear as the saddle of free energy (corresponding to the local maxima), although they cannot be captured in the canonical treatment.
Despite the presence of the thermodynamic phases being known, it has not been clarified how to describe them in the language of QFTs.
This has arisen the concept of partial deconfinement~\cite{Hanada:2016pwv,Berenstein:2018lrm,Hanada:2018zxn} as the counterpart of them.

%-------
\section{Application to deconfinement in SU(3) QCD}
\label{sec:QCD}
%-------
The basic idea for generalizing the notion of the large-$N$ deconfinement is to properly utilize the spatial extent of finite-$N$ QFTs.
The Polyakov line $P_{\vec{x}}$ defined at each spatial point $\vec{x}$ enables us to construct an analogous formulation to the large-$N$ theories. 
The set of $P_{\vec{x}}$ is identified with an element in a huge group $G = \prod_{\vec{x}}[{\rm SU}(N)]_{\vec{x}}$. 
There are $N$ eigenvalues at each point and $N V$ eigenvalues with the spatial volume $V$. It becomes sufficiently large in the thermodynamic limit $V\to \infty$ instead of the large-$N$ limit in this case, and this makes the distribution of the Polyakov line phases a continuous function. 
Hence, an analogous enhancement discussed in the previous section is expected for the partition function due to the infinite-volume limit even if the large-$N$ limit is not taken.

In this work, we employ lattice QCD computation to study the phase structure of finite-$N$ theories nonperturbatively.
The configurations for thermal QCD are generated by WHOT-QCD collaboration~\cite{Umeda:2012er}, in which $O(a)$-improved Wilson quark action~\cite{Sheikholeslami:1985ij} and the RG-improved Iwasaki gauge action~\cite{Iwasaki:1983iya} was used to simulate $N_{\rm f}=2+1$ QCD. 
Lattice spacing is $a\simeq 0.07$ fm, up- and down-quark mass heavier than physical mass ($m_\pi/m_\rho\simeq 0.63$) and almost physical strange-quark mass. 
For the scale setting, the $\rho$-meson mass was used as an input parameter.

In this section, we consider bare, rather than renormalized, Polyakov loops.
We will give a brief discussion about the issue of renormalization in the final section.
Note also that we show only the real part of $\tilde{\rho}_n$'s because they are real from the theoretical point of view. 
We confirmed that the imaginary part is consistent with zero.

In this setup, the multiply-wound Polyakov loops are slightly modified as 
\begin{equation}
    \tilde{\rho}_n = \frac{1}{N}\ev{\Tr (P_{\vec{x}})^n},
\end{equation}
where the bracket represents the spatial average. 
We can obtain the eigenphase distribution of the Polyakov line by replacing $u_n$ to $\tilde{\rho}_n$ in \eqref{eq:rho_Fourier_exp}.
Note that the distribution is defined as a continuous function due to the infinite-volume limit, similar to the large-$N$ theories.

At low temperatures, some ``uniform'' distribution $\rho(g)$ on $G$ would be realized, which reflects the symmetry of the confined vacuum in the language of the extended Hilbert space and the slowly varying nature of the Polyakov lines.
Such a distribution is the Haar-random distribution.
The uniform distribution in the confined phase at large $N$, $\rho_\mathrm{conf}=\frac{1}{2\pi}$ is indeed the ``SU$(\infty)$'' Haar-random distribution.
Then, the SU(3) Haar-random distribution\footnote{
For arbitrary $N$ and its generalization, see also Refs.~\cite{Hanada:2023rlk-ptep,Nishigaki:2024phx}. 
} is given by 
\begin{equation}
    \rho_\mathrm{Haar}(\theta)
    =
    \frac{1}{2\pi}\qty(
        1+\frac{2}{3}\cos(3\theta)
    ).
\end{equation}

\begin{figure}[t]
    \centering
    \scalebox{0.3}{\includegraphics{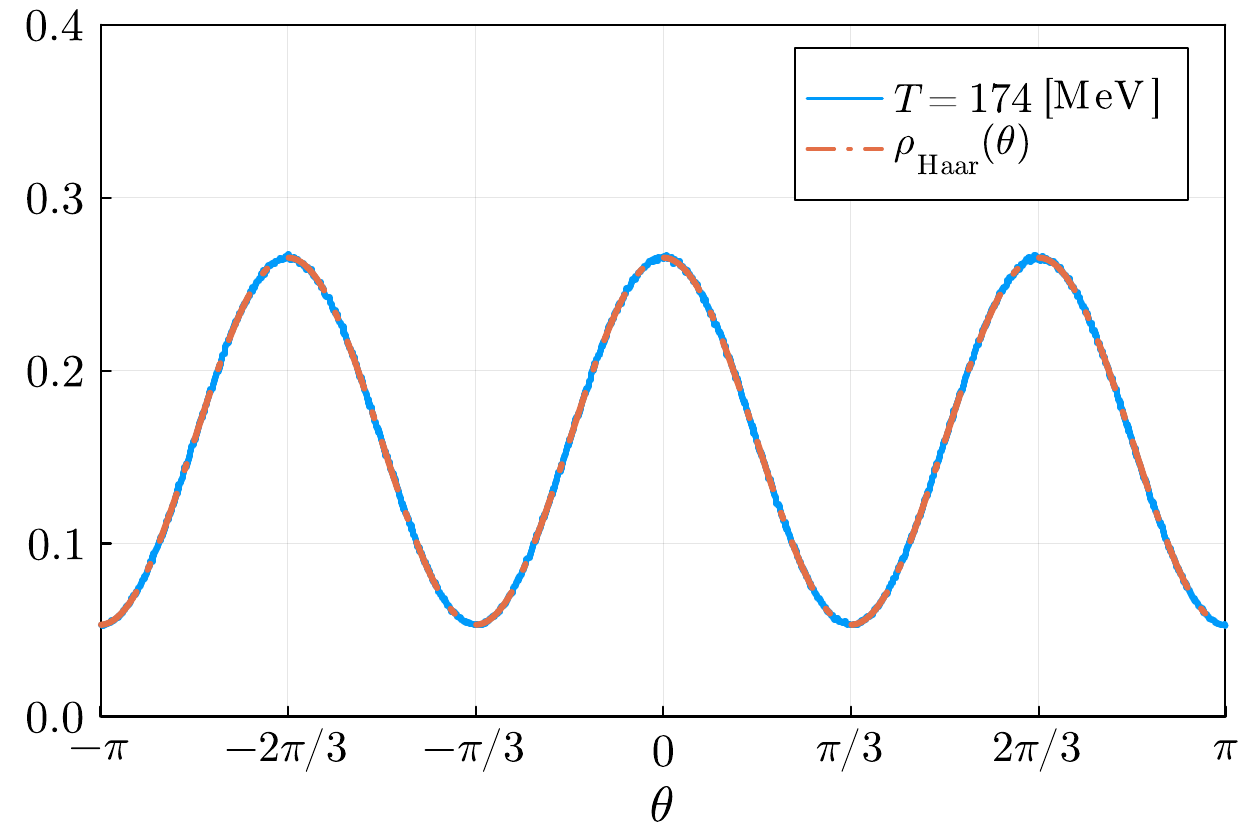}}
    \scalebox{0.3}{\includegraphics{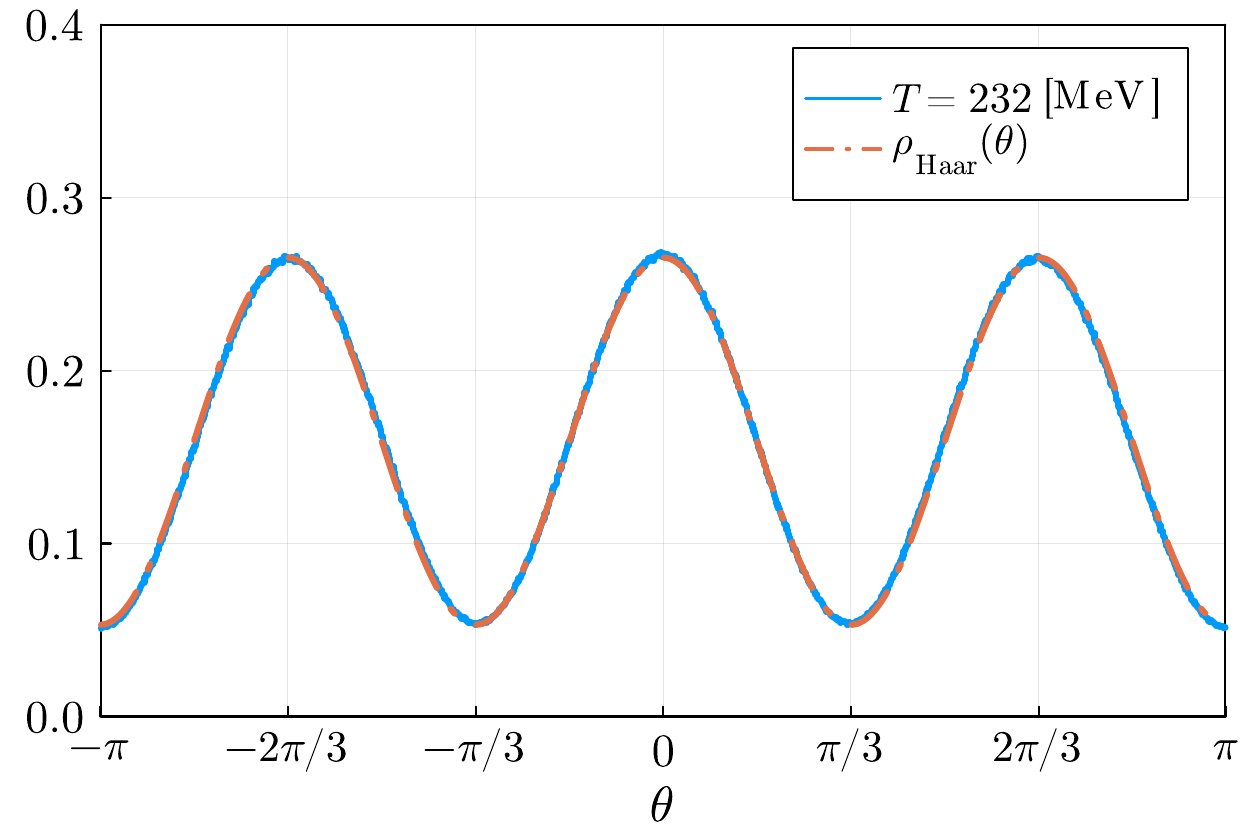}}
    \scalebox{0.3}{\includegraphics{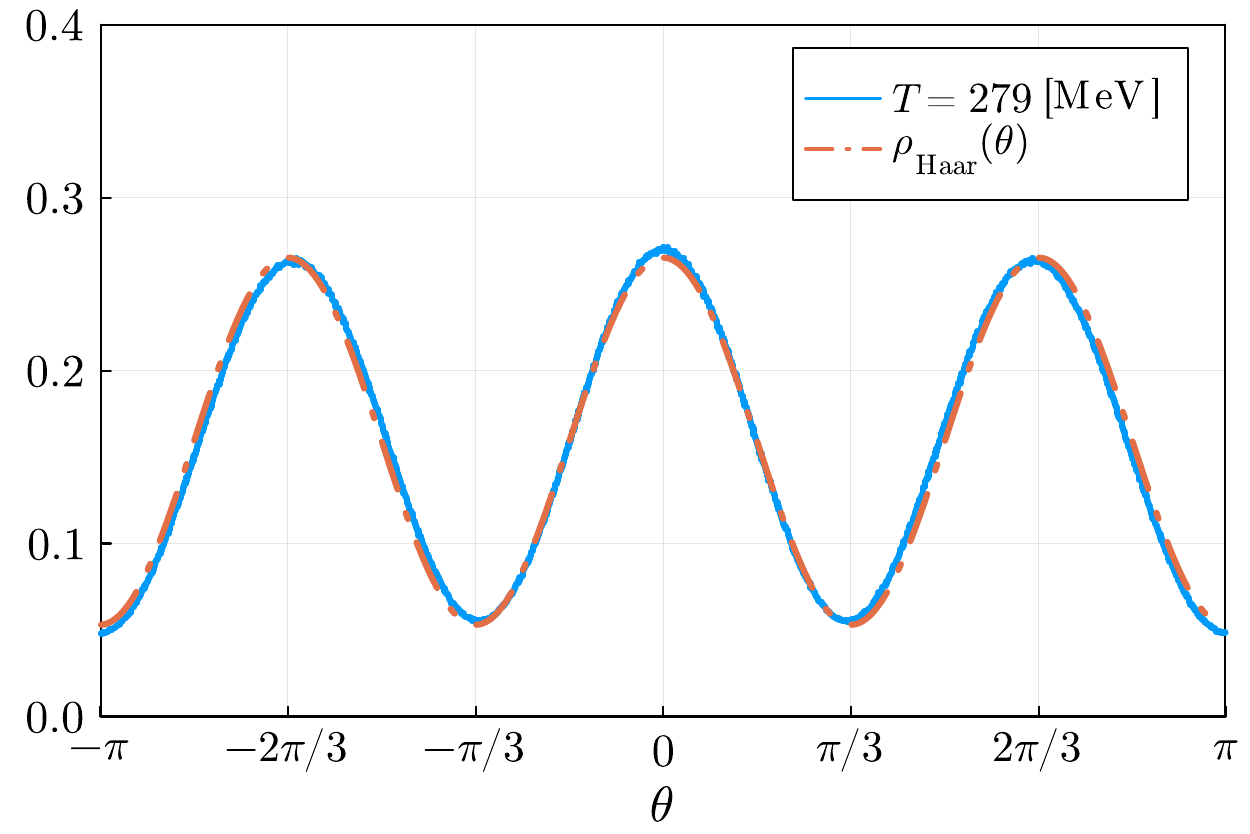}}
    \scalebox{0.3}{\includegraphics{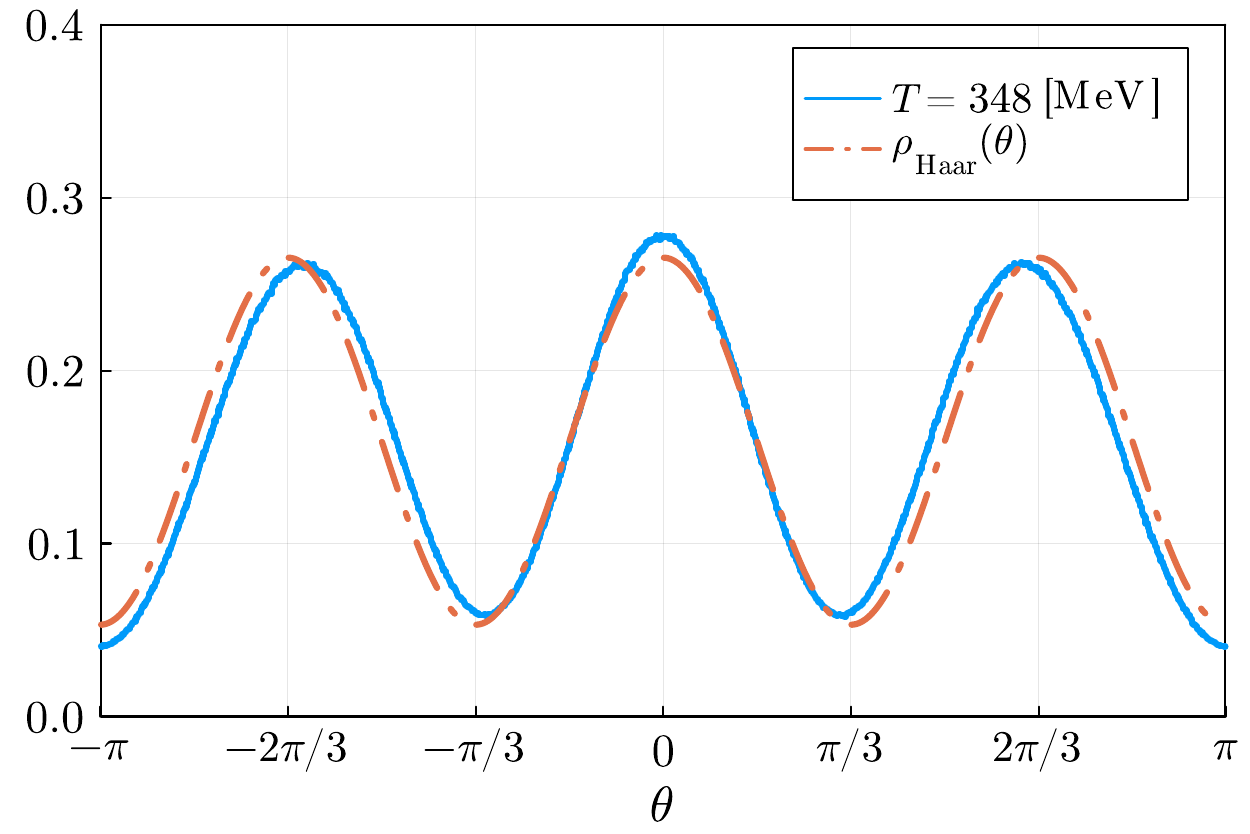}}
    \caption{Plots of the distribution of the Polyakov line for several temperatures $T=174, 232, 279, 348$ [MeV] (correspondingly, $N_t = 16, 12, 10, 8$), evaluated on the gauge configurations by the WHOT-QCD collaboration~\cite{Umeda:2012er}. The orange dashed lines represent the Haar-random distribution $\rho_\mathrm{Haar}(\theta)$, showing that the deviation between $\rho(\theta)$ and $\rho_\mathrm{Haar}(\theta)$ grows at higher temperatures. 
    }
    \label{fig:phase-dist_whot}
\end{figure}

\begin{figure}[t]
    \centering
    \includegraphics[width=\textwidth]{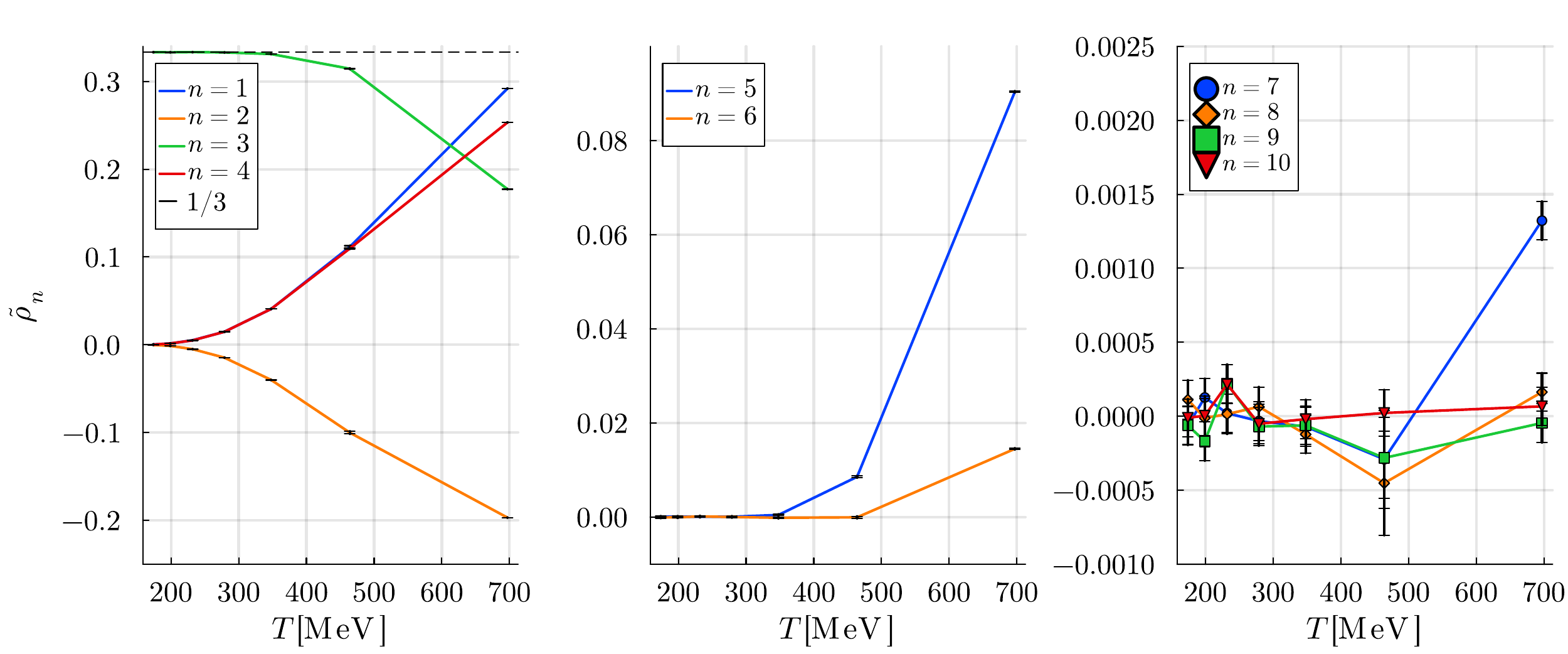}
    \\
    \scalebox{0.3}{\includegraphics{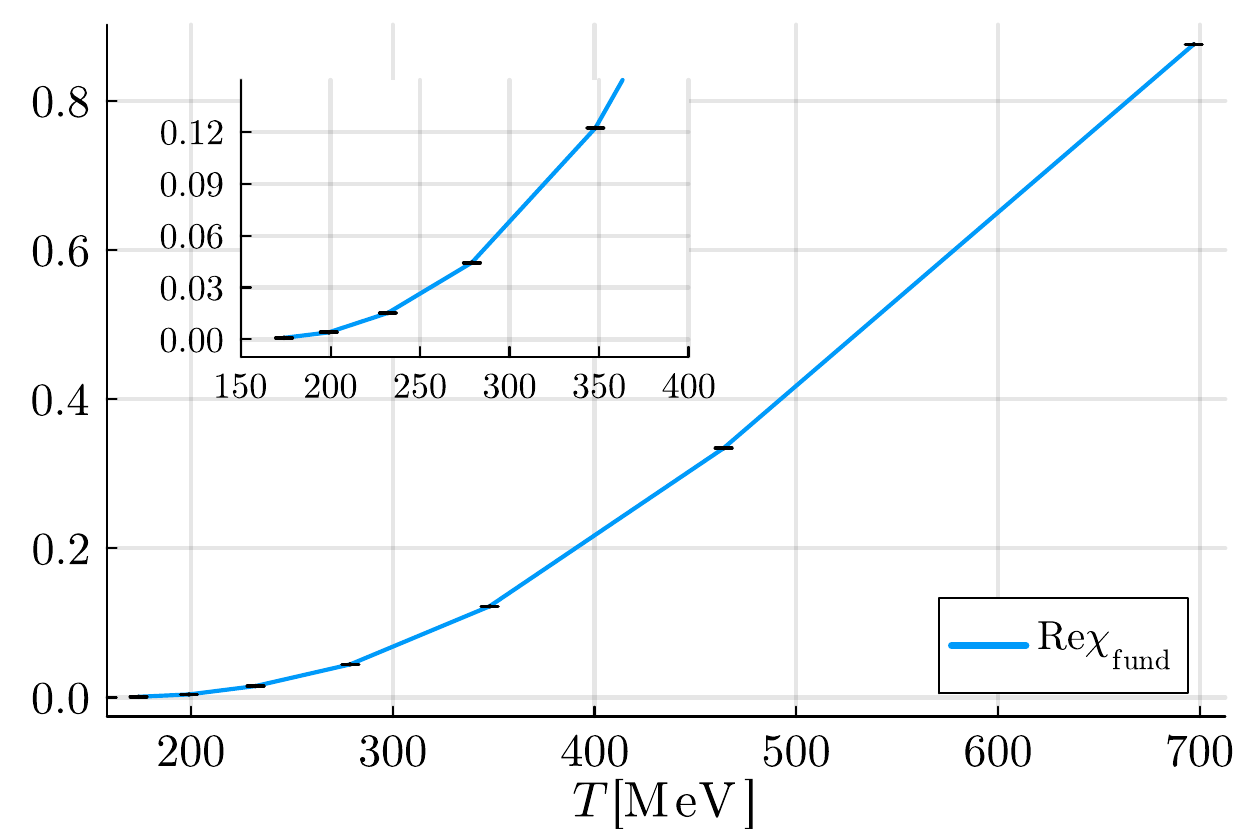}}
    \scalebox{0.3}{\includegraphics{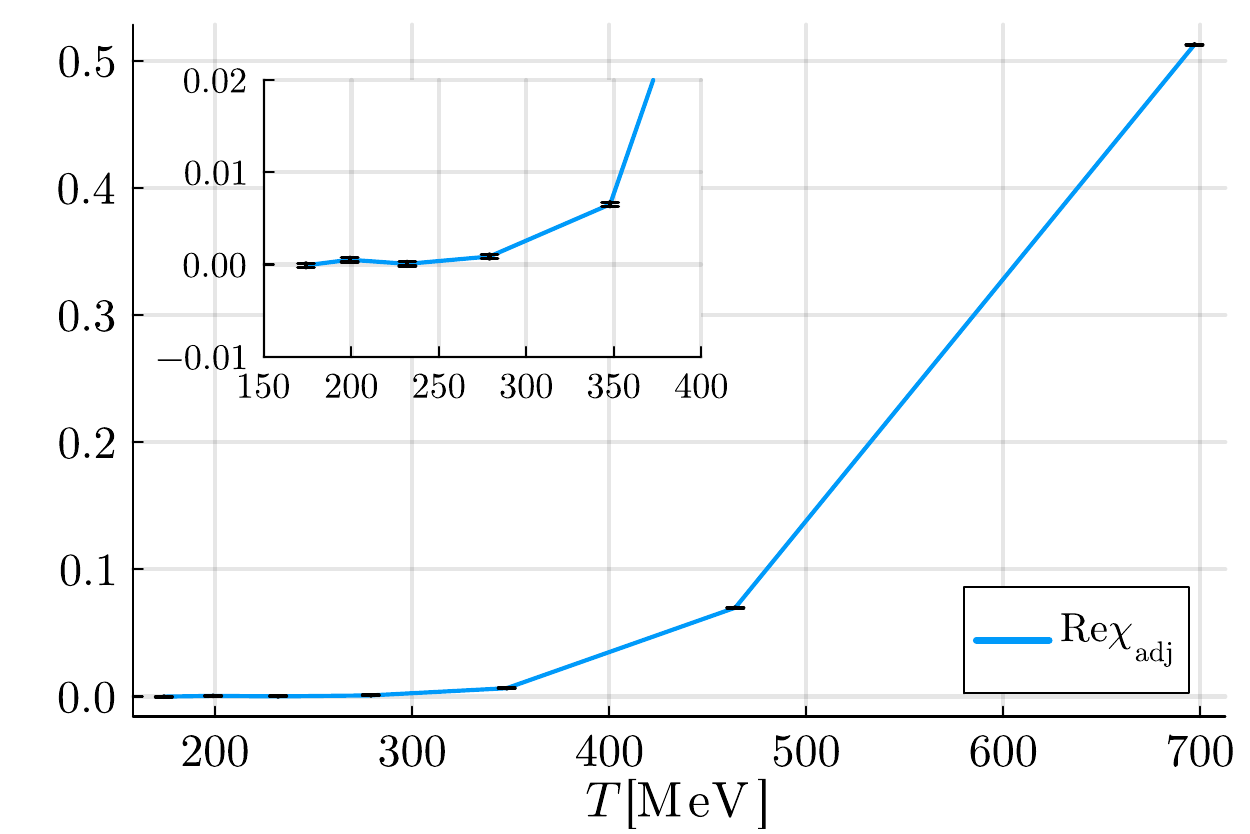}}
    \scalebox{0.3}{\includegraphics{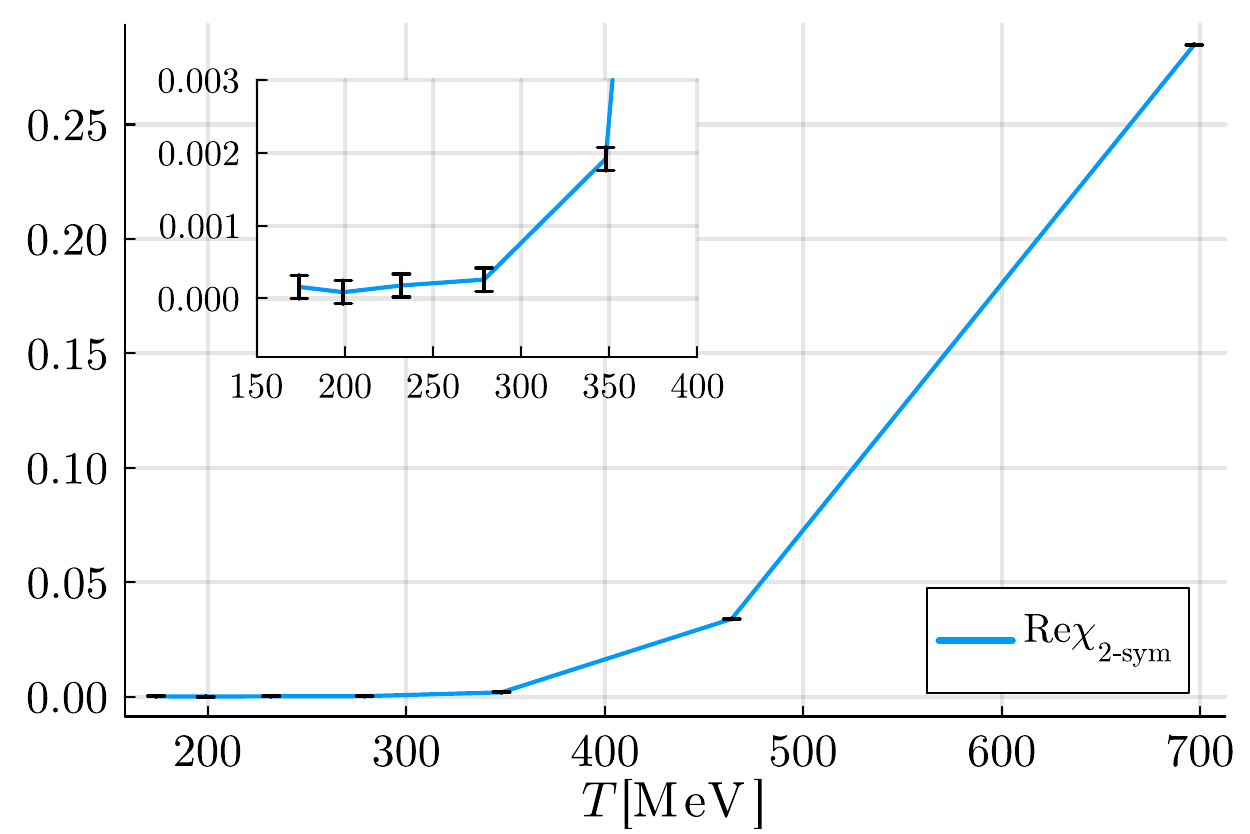}}
    \scalebox{0.3}{\includegraphics{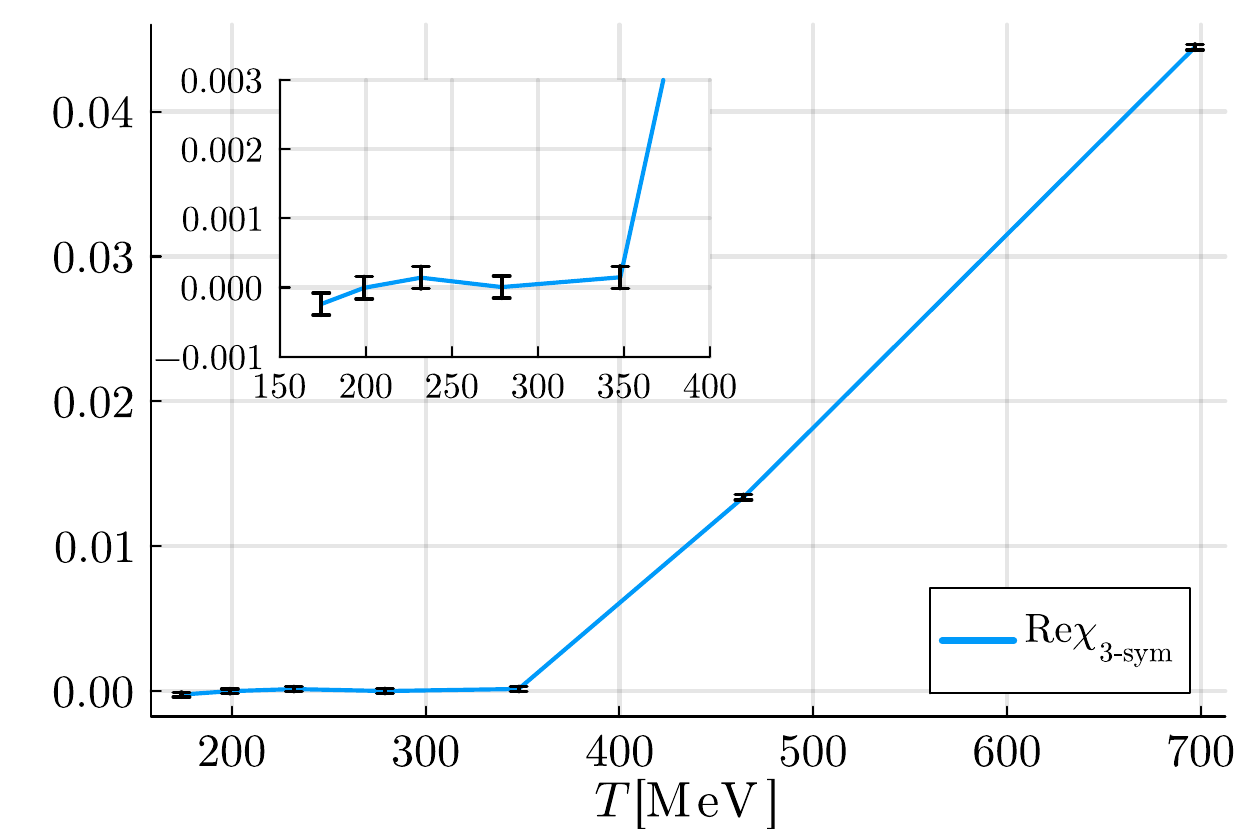}}
    \caption{
    Plots of the expectation values evaluated on the gauge configurations by the WHOT-QCD collaboration~\cite{Umeda:2012er}.
    [Top] multiply-wound Polyakov loops with winding $n = 1,\cdots,10$.  
    [Middle and Bottom] characters in the fundamental, adjoint, rank-2 symmetric, and rank-3 symmetric representations.
    }
    \label{fig:characters_whot}
\end{figure}

The phase distributions evaluated by the lattice configuration at different temperatures are shown in Fig.~\ref{fig:phase-dist_whot}.
A remarkable coincidence with $\rho_\mathrm{Haar}(\theta)$ can be observed at low temperatures, and the deviation from that distribution becomes clearer as the temperature is raised. 
The spatial average of the multiply-wound Polyakov loops with $n = 1,\cdots, 10$ is plotted at the top row of Fig.~\ref{fig:characters_whot}.
Their behaviors at low temperatures are consistent with the prediction by the Haar randomness, namely, $\tilde{\rho}_{n\ne 3} = 0$ and $\tilde{\rho}_3 = \frac{1}{3}$, up to numerical accuracy.
The temperature departing from the value expected by the Haar random is different for the modes.
Hence, we speculate that $T \ge 199$ MeV should be the partially deconfined phase.

In order for the physical interpretation to be more transparent, there is a better description than that with the multiply-wound Polyakov loops $\tilde{\rho}$, the use of \textit{Polyakov loops in the various representations}, or more precisely, the characters of the gauge group $G$.
For $g \in G$, the character in the representation $r$ can be defined by the trace of the corresponding representation matrix $R_r(g)$ as
\begin{equation}
    \chi_r(g)
    =
    \tr R_r(g).
\end{equation}
The reason that characters form a convenient basis on the gauge group $G$ is that the characters in the irreducible representations $r, r'$ satisfy the orthonormal condition
\begin{equation}
    \frac{1}{\Vol(G)}
    \int \dd g\, \chi_r(g)\chi^\ast_{r'}(g) = 1.
\end{equation}
Since the gauge element $g$ is now holonomy $P$, the Polyakov loops are the functions of the eigenphases. Specifically, for $G=$ SU(3), 
\begin{equation}
    \chi_r =
    \begin{cases}
        %\displaystyle
        \sum_{j=1}^3 \ee^{\ii\theta_j}      &       (r:\textrm{fundamental})
        \\
        %\displaystyle
        2 + \sum_{j\neq k} \ee^{\ii(\theta_j-\theta_k)}     &       (r:\textrm{adjoint})
        \\
        %\displaystyle
        \sum_{j=1}^3 \ee^{2\ii\theta_j}
        +
        \sum_{j=1}^3 \ee^{-\ii\theta_j}        &        (r:\textrm{rank$\mathchar`-2$ symmetric})
        \\
        %\displaystyle
        1 + \sum_{j=1}^3 \ee^{3\ii\theta_j} 
        + 
        \sum_{j\neq k} \ee^{\ii(\theta_j-\theta_k)}     &       (r:\textrm{rank$\mathchar`-3$ symmetric})
    \end{cases}
    ,
\end{equation}
and so on.
The multiply-wound loops are also functions of the holonomy, and therefore, they can be expressed in terms of the characters;
\begin{align}
    3 u_1 &= \chi_{\rm fund.}\, ,
    \label{eq:u1_by_characters}
    \\
    3 u_2 
    &= 
    \chi_{\rm 2\mathchar`-sym.} - \left(\chi_{\rm fund.}\right)^\ast\, ,
    \\
    3 u_3
    &=
    1 + \chi_{\rm 3\mathchar`-sym.} - \chi_{\rm adj.}\, ,
    \label{eq:u3_by_characters}
\end{align}
and so on.
When $P$ is Haar random, $\ev{\chi_r} = 0$ for any nontrivial representation. 
Notice also that the nonzero excitation values imply the deconfinement of the excitation in the corresponding representation.

The middle and bottom rows of Fig.~\ref{fig:characters_whot} plot the expectation values of characters in various representations.
These plots show that $\langle\chi_{\rm fund.}\rangle$ becomes nonzero at around $174~{\rm MeV}$ and $\langle\chi_{\rm adj.}\rangle$, $\langle\chi_{\rm 2\mathchar`-sym.}\rangle$, $\langle\chi_{\rm 3\mathchar`-sym.}\rangle$ depart from zero at higher temperatures ($T\gtrsim 348$ MeV), where the $n$-wound loops with $n=5,6$ set in. 
It would be supporting evidence to interpret that the onset of $\langle\chi_{\rm adj.}\rangle$, $\langle\chi_{\rm 2\mathchar`-sym.}\rangle$ and $\langle\chi_{\rm 3\mathchar`-sym.}\rangle$ is accompanied with the transition to separate the nontrivial intermediate phase. 
We emphasize that the behaviors of the Polyakov loops in the nontrivial representations explain completely those of the multiply-wound loops.
This may indicate the validity of the classical relations \eqref{eq:u1_by_characters}-\eqref{eq:u3_by_characters} even at the quantum level.
It is important to examine whether these relations hold when taking the continuum limit with an appropriate renormalization procedure.

\begin{figure}[th]
    \centering
    \scalebox{0.5}{\includegraphics{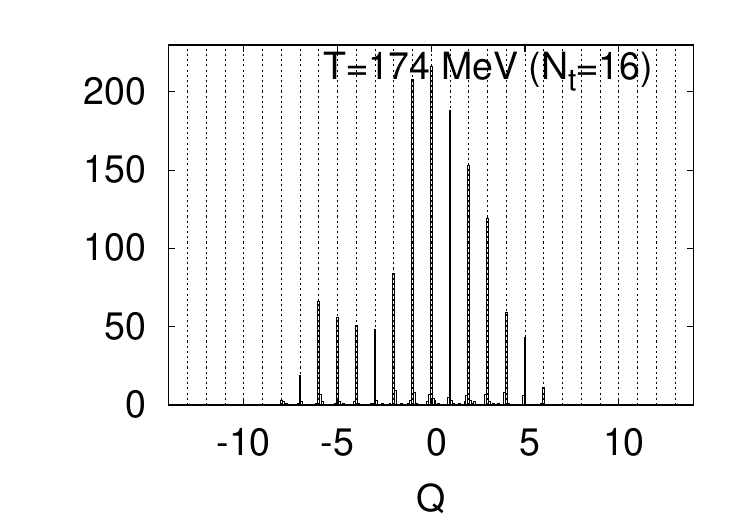}}
    \scalebox{0.5}{\includegraphics{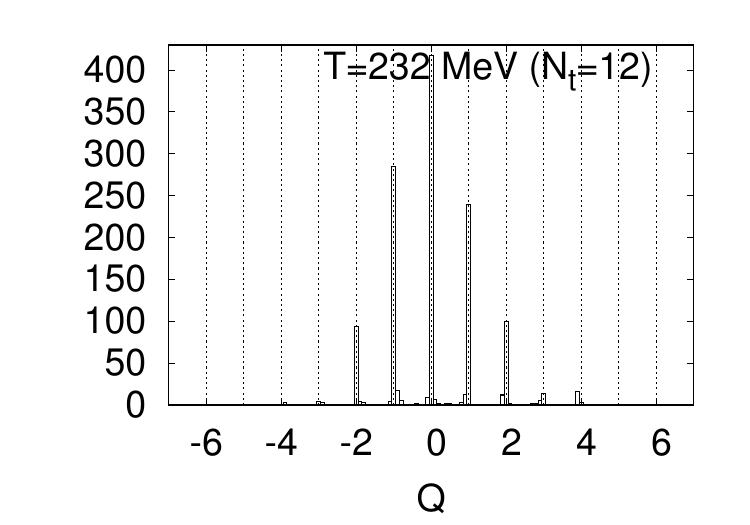}}
    \scalebox{0.5}{\includegraphics{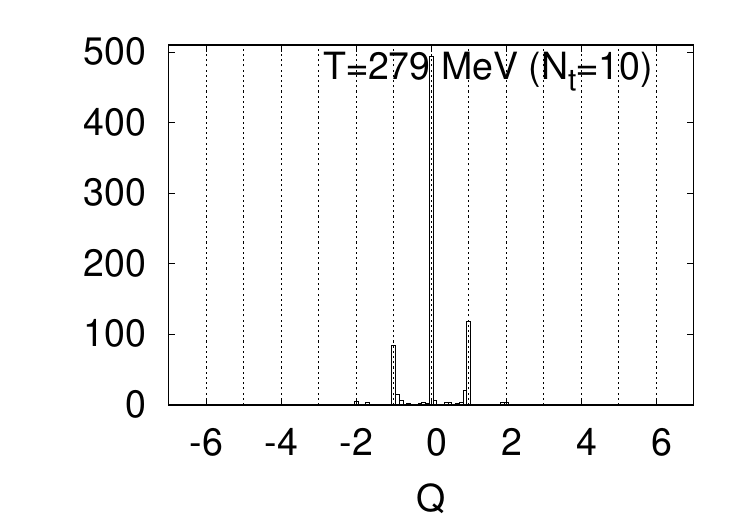}}
    \scalebox{0.5}{\includegraphics{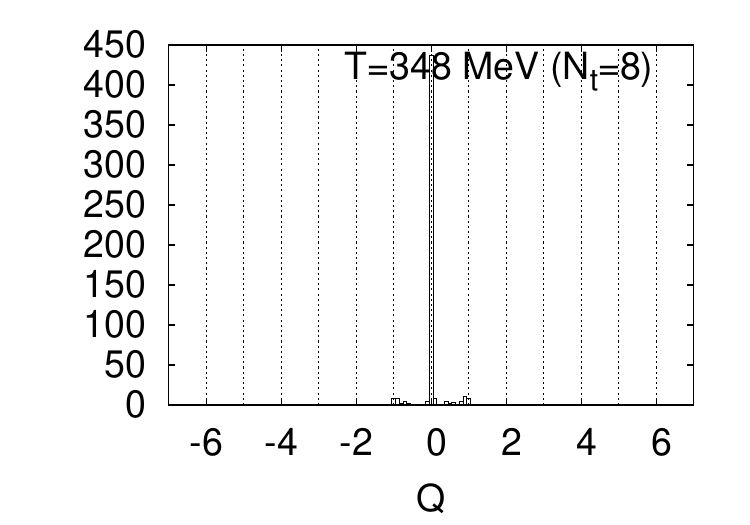}}
    \caption{Histograms of the topological charge for several temperatures $T=174, 232, 279, 348$ [MeV] (correspondingly, $N_t = 16, 12, 10, 8$), provided by the WHOT-QCD collaboration~\cite{Taniguchi:2016tjc,Umeda:2012er}. They are obtained through the gradient flow.
    }
    \label{fig:Qtop_whot}
\end{figure}

We further discuss that the higher ``transition'' temperature has a certain physical interpretation from the traditional viewpoint of QCD. 
The finite-$N$ counterpart of the GWW transition is expected to be characterized by the global symmetry of QCD. 
Chiral symmetry is an approximate symmetry of QCD since quarks are massive in the Universe, and the correspondence~\cite{Hanada:2021ksu} between the chiral symmetry breaking/restoration for the massless probe quarks and the GWW transition discussed in the large-$N$ theory does not apply directly to the current situation.
Interestingly, we observe that the instantons, which are intimately connected by the chiral symmetry, condense around the same temperature scale and fit nicely to our criterion of phase characterization.
The instanton condensation was already measured by the WHOT-QCD collaboration~\cite{Taniguchi:2016tjc} through the topological charge of each lattice configuration.
By the smearing with, for example, the gradient flow~\cite{Luscher:2010iy}, each configuration gives the histogram of the topological charge that has peaks at integer values, as plotted in Fig.~\ref{fig:Qtop_whot}.
The plots show multiple clear peaks, including the ones at $Q \neq 0$, below $ T = 279$ MeV. 
Wider distributions can be observed at lower temperatures.
On the other hand, the plots higher than the temperature have a single peak at $Q = 0$, indicating the occurrence of the instanton condensation.
Therefore, we claim that there is a phase transition around $T\lesssim$ 348 MeV that correlates with the behaviors of the excitation modes in the higher representations.

%-------
\section{Conclusion and discussion}
\label{sec:conclusion}
%-------
We discussed the application of the notion of the deconfinement phase transition cultivated at large-$N$ to the finite-$N$ gauge theories, in particular, SU(3) QCD with fermions.
From the numerical examinations using the lattice QCD gauge configurations produced by the WHOT-QCD collaborations, we proposed that the Polyakov loops in the various representations are useful quantities to characterize the thermal phases in gauge theories including QCD.
In this work, we found that the Haar-random distribution plays a crucial role in the confining vacuum for both large-$N$ and finite-$N$ theories. The deviation, which becomes clearer as the temperature is higher, can be measured qualitatively by the Polyakov loops in the nontrivial representations. In the large-$N$ theories, the GWW transition specifies the border between partially- and completely-deconfined phases, and we discussed that the analog of the transition for QCD is observed as the instanton condensation by the heuristic argument based on the numerical analysis. It is worthwhile to clarify the relationship between the instanton condensation and the global symmetry of QCD in a more detailed manner.

To characterize the thermodynamic phases, we considered the bare Polyakov lines, not renormalized ones, since the former is responsible for the symmetry in the extended Hilbert states, as discussed in Sec.~\ref{sec:mechanism}.
As far as we deal with the theory defined with a UV cutoff for different temperatures, our characterization is well-defined without the help of renormalization.
From this stance, we employed the Polyakov loops in various representations to characterize the phases. 
However, the issue of renormalization is inevitable when comparing different cutoff theories and, furthermore, taking the continuum limit. 
The proper renormalization schemes should possess multiplicative renormalizability without mixing and preserve the correct behavior when deviating from the Haar-random distribution. Specifying suitable schemes is an important future prospect.

Returning to the original viewpoint and considering the application of the characters to quantum gravity is also worthwhile. 
The finite-$N$ nature of the large-$N$ deconfinement will help us understand the quantum aspects of black hole physics.
The technique of character expansion then becomes a powerful tool for the purpose, as it has been in the past~\cite{Gross:1993hu,Gross:1993yt,Berenstein:2023srv}.

\acknowledgments
We would like to thank M. Hanada, H. Ohata, and H. Shimada for collaborations~\cite{Hanada:2023krw-ptep,Hanada:2023rlk-ptep} produced the results in this proceeding. 
We also would like to thank the members of the WHOT-QCD collaboration, including S. Ejiri,
K. Kanaya, M. Kitazawa, and T. Umeda, for providing us with their lattice configurations and many
plots and having stimulating discussions with us. The analysis of topological charge in Ref.~\cite{Taniguchi:2016tjc},
which was crucial in Sec.~\ref{sec:QCD}, was led by Y. Taniguchi who passed away in 2022. K. Kanaya collected the data and plots created by Y. Taniguchi for us. We deeply thank Y. Taniguchi and K. Kanaya. This work was supported by the Japan Lattice Data Grid (JLDG) constructed over the SINET5 of NII. H.W. was supported by
the Japan Society for the Promotion of Science (JSPS) KAKENHI Grant number 22H01218.

\bibliographystyle{JHEP}
\bibliography{ref}

\end{document}